\documentclass[conference]{IEEEtran}
\IEEEoverridecommandlockouts

\usepackage{cite}
\usepackage{amsmath,amssymb,amsfonts}
\usepackage{algorithm}
\usepackage{algorithmic}
\usepackage{graphicx}
\usepackage{textcomp}
\usepackage{xcolor}
\usepackage{booktabs}
\usepackage{tabularx}
\usepackage{array,multirow}
\makeatletter
\renewcommand{\footnoterule}{%
  \kern -3pt
  \hrule width \columnwidth height 0.4pt
  \kern 2.6pt
}
\makeatother

\def\BibTeX{{\rm B\kern-.05em{\sc i\kern-.025em b}\kern-.08em
    T\kern-.1667em\lower.7ex\hbox{E}\kern-.125emX}}
\begin{document}

\title{A Dynamic Prognostic Prediction Method for Colorectal Cancer Liver Metastasis}

\author{
\IEEEauthorblockN{
Wei Yang\textsuperscript{1,*},
Yiran Zhu\textsuperscript{1,*},
Yan Su\textsuperscript{2},
Zesheng Li\textsuperscript{2},
Chengchang Pan\textsuperscript{2,\textdagger},
and Honggang Qi\textsuperscript{2,\textdagger}
}
\IEEEauthorblockA{
\textsuperscript{1}Department of Computer, North China Electric Power University, Baoding, China\\
Emails: wei\_yang\_edu@yeah.net, ciaran\_study@yeah.net
}
\IEEEauthorblockA{
\textsuperscript{2}School of Computer Science and Technology, University of the Chinese Academy of Sciences, Beijing, China\\
Emails: 1227698971@qq.com, 2022111515@stu.sufe.edu.cn, chpan.infante@qq.com, hgqi@ucas.ac.cn
}
\IEEEauthorblockA{
\textsuperscript{*}Equal contribution \qquad
\textsuperscript{\textdagger}Corresponding authors
}
}

\maketitle

\begin{abstract}
Colorectal cancer liver metastasis (CRLM) exhibits high postoperative recurrence and pronounced prognostic heterogeneity, challenging individualized management. Existing prognostic approaches often rely on static representations from a single postoperative snapshot, and fail to jointly capture tumor spatial distribution, longitudinal disease dynamics, and multimodal clinical information, limiting predictive accuracy. We propose DyPro, a deep learning framework that infers postoperative latent trajectories via residual dynamic evolution. Starting from an initial patient representation, DyPro generates a 12-step sequence of trajectory snapshots through autoregressive residual updates and integrates them to predict recurrence and survival outcomes. On the MSKCC CRLM dataset, DyPro achieves strong discrimination under repeated stratified 5-fold cross-validation, reaching a C-index of 0.755 for OS and 0.714 for DFS, with OS AUC@1y of 0.920 and OS IBS of 0.143. DyPro provides quantitative risk cues to support adjuvant therapy planning and follow-up scheduling.
\end{abstract}

\begin{IEEEkeywords}
Postoperative Management of CRLM, Personalized Prognostic Prediction, Dynamic Risk Assessment, Multimodal learning, Graph neural networks
\end{IEEEkeywords}

\section{Introduction}
Colorectal cancer liver metastasis (CRLM) is a major contributor to colorectal cancer--related mortality. Prior studies report that half of colorectal cancer patients develop liver metastases during the disease course \cite{birrer2021}. For selected patients with resectable disease, surgery combined with systemic therapy can achieve favorable long-term survival, yet postoperative recurrence remains common and prognosis is highly heterogeneous. Reliable preoperative risk stratification is therefore essential for individualized perioperative planning and postoperative surveillance.

Conventional prognostic tools largely rely on clinicopathologic variables and risk scores, offering limited personalization and showing variable robustness across cohorts \cite{fong1999,kokkinakis2024}. A key limitation is that they do not systematically leverage imaging evidence from routine preoperative CT, including lesion spatial distribution and spatial relationships to liver parenchyma and vascular structures. With advances in medical AI, deep models can learn prognostic representations from CT for recurrence and survival prediction. However, unimodal imaging alone cannot capture the multifactorial nature of tumor progression. Multimodal learning that jointly models imaging and key clinical indicators has shown promise, but many existing pipelines still adopt coarse fusion, which weakens modeling of intrinsic cross-modal interactions and limits generalization.

CRLM progression is also inherently dynamic. Prognostic signals are reflected in temporal patterns such as lesion growth and recurrence timing, while most existing models remain based on a single preoperative snapshot. Although longitudinal imaging studies suggest that temporal information is valuable for outcome modeling \cite{qu2023,jin2021}, retrospective surgical cohorts often contain sparse and irregular follow-up, making it difficult to directly learn temporal models from observed sequences. This motivates latent trajectory inference: a model can infer a plausible postoperative risk trajectory via autoregressive residual state transitions, without requiring dense longitudinal observations.

\begin{figure*}[t]
  \centering
  \includegraphics[trim=0 220pt 0 360pt, clip, width=0.94\linewidth]{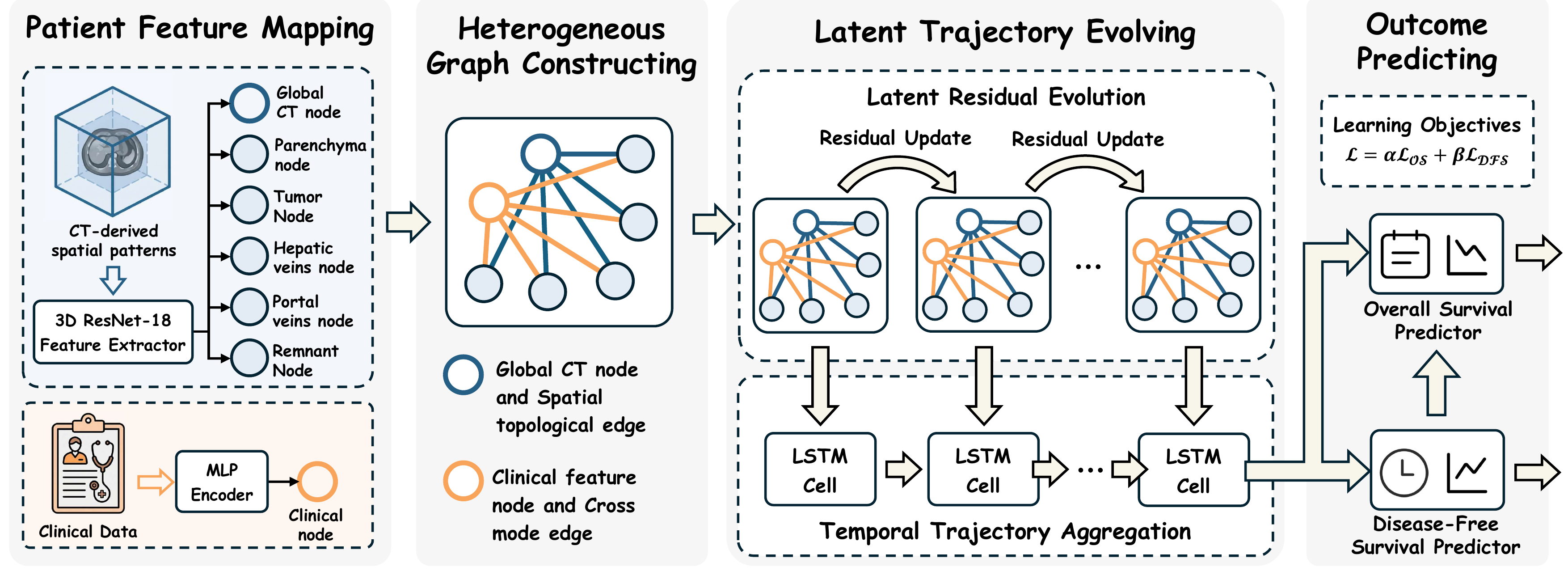}
  \caption{Overview of DyPro. Preoperative CT and clinical data are mapped to a heterogeneous patient graph, evolved by latent residual dynamics, and aggregated by an LSTM to jointly predict DFS and OS.}
  \label{fig2}
\end{figure*}

To address these gaps, we propose DyPro for individualized prediction of postoperative recurrence and survival in CRLM. DyPro integrates CT-derived spatial patterns with key clinical indicators and infers postoperative disease dynamics as a latent trajectory via autoregressive state transitions. It uses residual evolution to generate long-horizon trajectory snapshots and a sequence integrator to aggregate progression cues for outcome prediction. We evaluate DyPro on a publicly released MSKCC CRLM dataset pairing preoperative hepatic CT with clinicopathologic variables and recurrence/survival endpoints \cite{simpson2024}.

Compared to existing methods, our main contributions are summarized as follows:
\begin{itemize}
    \item We propose DyPro, a multimodal prognostic framework for CRLM that encodes preoperative CT--derived spatial patterns and key clinical indicators in a heterogeneous patient graph to enable individualized prediction of postoperative DFS and OS.
    \item To address the limitation of static preoperative models that ignore postoperative progression, we design a dynamic latent residual evolution module that rolls the patient graph forward via autoregressive state transitions and aggregates the resulting trajectory for postoperative DFS and OS prediction.
    \item Extensive experiments on the MSKCC CRLM cohort demonstrate that DyPro achieves strong and robust recurrence and survival-time prediction, consistently improving discrimination and calibration over competitive clinical and radiomics baselines.
\end{itemize}

\section{Related Work}
Early prognostic studies for colorectal liver metastases (CRLM) mainly relied on clinicopathologic variables and risk scores. Fong et al. proposed a clinical score for recurrence risk stratification using readily available criteria, but its linear assumptions limit expressivity \cite{fong1999}. Regression-based survival modeling, including Cox-type formulations \cite{andersen1982}, remains common, yet typically depends on a small set of hand-crafted variables and does not systematically leverage imaging evidence. With advances in medical imaging analytics, radiomics and deep learning have been used to learn prognostic representations from preoperative CT. Liu et al. adopted a 3D CNN architecture to capture spatial heterogeneity and improve outcome prediction \cite{liu2022}. However, most CT-only approaches remain snapshot-based and cannot explicitly integrate multimodal factors or represent postoperative risk dynamics.

To overcome the limitations of unimodal models, multimodal fusion combines imaging with patient information. Kickingereder et al. integrated imaging with clinical and genetic factors for improved survival prediction \cite{kickingereder2016}. Subsequent deep multimodal systems further demonstrated the benefit of joint modeling in oncology applications \cite{mckinney2020,zhou2023}. However, many methods still rely on coarse fusion such as late fusion or direct concatenation, which under-models cross-modal interactions and can limit generalization across cohorts.

Graph neural networks (GNNs) provide a principled approach for modeling non-Euclidean structures and structured relationships in medical data \cite{gogoshin2023}. Fu et al. constructed a multimodal graph network combining imaging-derived phenotypes with patient variables to refine risk stratification \cite{fu2023}. Wang et al. designed a dual-stream graph framework to model complementary signals from different modalities \cite{wang2024}, and Yang et al. integrated similarity networks to extract higher-order prognostic features \cite{yang2024}. While effective, these methods typically treat patient data as static graphs and overlook temporal disease dynamics that are critical for recurrence and survival prediction in CRLM.

Temporal information is clinically relevant for prognosis, and longitudinal imaging analyses suggest that multi-timepoint evidence can provide additional predictive value. Qu et al. introduced dynamic radiomics for CRLM treatment efficacy prediction and reported improvements over conventional radiomics \cite{qu2023}. Jin et al. explored learning from longitudinal images for treatment response prediction \cite{jin2021}. However, retrospective surgical cohorts often contain sparse and irregular follow-up, making it difficult to directly learn temporal models from observed sequences. Spatiotemporal graph learning in computer vision, such as ST-GCN \cite{yan2018}, demonstrates the potential of jointly modeling relations and evolution, but direct adoption in clinical prognosis is non-trivial when longitudinal observations are limited. Motivated by these gaps, our work aims to strengthen multimodal interaction modeling while representing postoperative dynamics via latent trajectory inference for individualized CRLM prognosis prediction.

\section{Method}
As shown in Fig.~\ref{fig2}, DyPro comprises patient-specific tumor mapping, postoperative dynamic risk tracking, and prognostic heads with learning objectives.

\subsection{Patient-Specific Tumor Mapping Engine}
As shown in Fig.~\ref{fig2}, we represent each patient as a heterogeneous graph constructed from preoperative CT and clinical variables.
From contrast-enhanced CT, we segment five anatomical/disease regions as image nodes: liver parenchyma, future liver remnant, hepatic veins, portal veins, and metastatic tumors; we additionally include a global CT node to encode overall hepatic context.
Each image node is encoded by a pretrained 3D-ResNet18 with multi-scale pooling to fuse local details and global structure.
All clinical indicators are standardized and mapped to a single clinical node embedding in the same latent space as image nodes.

Edges are defined by clinically interpretable rules.
The global CT node connects to each anatomical node via spatial topology edges whose attributes encode normalized 3D centroid offsets.
The clinical node connects to each anatomical node to associate clinical factors with region-specific imaging phenotypes.
In our design, each patient graph instantiates seven semantic nodes: five anatomical/disease region nodes, one global CT node, and one clinical node; the same construction principle can be extended to accommodate missing entities or additional regions when available.
This yields a clinically interpretable heterogeneous graph, serving as the input to the postoperative dynamic risk tracker.

\begin{algorithm}[t]
\caption{Latent Residual Evolution}
\label{alg:evolution}
\textbf{Input:} Initial node embeddings $H^{(0)} \in \mathbb{R}^{|V|\times d}$, time horizon $T$ \\
\textbf{Output:} State trajectory $H_{\mathrm{seq}}=\{H^{(t)}\}_{t=0}^{T-1}$ and snapshots $Z_{\mathrm{seq}}=\{z_t\}_{t=0}^{T-1}$
\begin{algorithmic}[1]
\STATE $H_{\mathrm{seq}} \leftarrow [\,H^{(0)}\,]$
\STATE $H \leftarrow H^{(0)}$
\FOR{$t = 0$ \TO $T-1$}
    \STATE $e_t \leftarrow \mathrm{Emb}(t)$ \quad // learnable time embedding
    \STATE $\Delta \leftarrow \mathcal{F}_{\theta}\big([H; e_t], E\big)$ \quad // GNN-based incremental change
    \STATE $H \leftarrow H + \Delta$ \quad // residual update
    \STATE $\mathrm{append}(H_{\mathrm{seq}}, H)$
    \STATE $z_t \leftarrow \mathrm{READOUT}(H)$ \quad // e.g., global mean pooling
    \STATE $\mathrm{append}(Z_{\mathrm{seq}}, z_t)$
\ENDFOR
\STATE \textbf{return} $H_{\mathrm{seq}}, Z_{\mathrm{seq}}$
\end{algorithmic}
\end{algorithm}

\subsection{Post-Op Dynamic Risk Tracker}
As shown in Fig.~\ref{fig2}, we propose a latent trajectory evolution module to model long-term postoperative risk dynamics.
Instead of reconstructing each follow-up horizon independently, DyPro treats the latent graph state as a discrete-time dynamical system: the state at year $t$ is obtained by updating the state at $t-1$ with a time-conditioned residual graph operator.

\textbf{Latent Residual Evolution.}
Let $H^{(t)} \in \mathbb{R}^{|V|\times d}$ denote the node representations on the patient graph at year $t$.
We introduce a learnable time embedding $e_t = \mathrm{Emb}(t) \in \mathbb{R}^{d_t}$ and evolve the state by
\begin{equation}
H^{(t)} = H^{(t-1)} + \mathcal{F}_{\theta}\!\big([H^{(t-1)}; e_t], E\big), \quad t = 1,\dots,T-1,
\label{eq:evolution}
\end{equation}
where $E$ denotes the fixed graph connectivity and $\mathcal{F}_{\theta}$ is a GNN-based residual operator.
In practice, we concatenate the time embedding $e_t$ to each node feature and apply a lightweight message-passing stack (GraphSAGE, GAT, or GCN depending on the chosen backbone) to obtain the incremental change between adjacent years.
This design explicitly couples temporal evolution with spatial tumor–liver context on the graph.

\textbf{Trajectory Readout.}
At each step, we obtain a graph-level snapshot by pooling node states:
\begin{equation}
z_t = \mathrm{READOUT}(H^{(t)}) = \mathrm{GlobalMeanPool}(H^{(t)}),
\end{equation}
which yields a latent trajectory sequence $\{z_t\}_{t=0}^{T-1}$.
This sequence is subsequently fed into a trajectory aggregator to produce a compact trajectory-aware representation for the downstream prognostic heads.
The overall residual evolution procedure is summarized in Algorithm~\ref{alg:evolution}.

\subsection{Prognostic Heads and Learning Objectives}
\label{sec:heads_objectives}

\textbf{Temporal trajectory aggregation.}
Given the trajectory snapshots $\{z_t\}_{t=0}^{N-1}$ from the dynamic risk tracker, we employ an LSTM as a trajectory integrator to aggregate long-term progression signals.
To avoid relying on a single time step, we aggregate hidden states across the horizon (e.g., mean pooling) to obtain a trajectory representation $h^{*}$ for prediction.

\textbf{Prediction heads.}
On top of $h^{*}$, we attach two discrete-time survival heads for disease-free survival (DFS) and overall survival (OS).
Each head outputs hazard logits over $K$ time bins, which are transformed into survival curves and then into predicted event times.
To capture the dependency between recurrence and death, we adopt a cascaded design: the DFS branch first projects $h^{*}$ to a DFS-specific context vector, and the OS head takes the concatenation of $h^{*}$ and this context as input, implementing DFS$\rightarrow$OS conditioning at the representation level.

\textbf{Loss function design.}
We train DyPro with standard discrete-time survival objectives.
Let $L_{\mathrm{OS}}$ and $L_{\mathrm{DFS}}$ denote the negative log-likelihood losses for OS and DFS, computed from the predicted hazards and the observed times/censoring indicators.
The overall training objective is
\begin{equation}
L = \alpha L_{\mathrm{OS}} + \beta L_{\mathrm{DFS}},
\end{equation}
where $\alpha$ and $\beta$ balance the two tasks.

\begin{figure}[t]
  \centering
  \includegraphics[width=0.9\columnwidth]{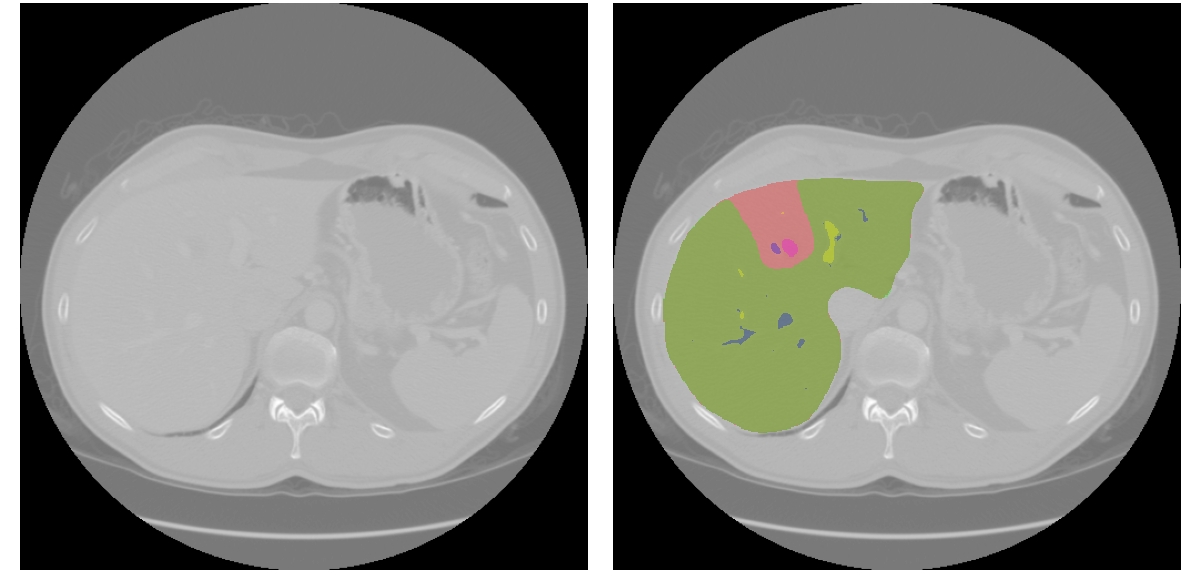}
  \caption{Visualization of the dataset, including segmentation of the liver, residual liver, hepatic veins, portal veins, and tumors.}
  \label{fig4}
\end{figure}

\begin{table*}[t]
\caption{DFS and OS results for different GNN backbones (mean$\pm$std over repeated stratified 5-fold CV with 3 repeats).}
\label{tab:gnn_backbone_twocol}
\centering
\footnotesize
\setlength{\tabcolsep}{4pt}
\renewcommand{\arraystretch}{1.15}
\begin{tabularx}{\textwidth}{>{\centering\arraybackslash}m{1.0cm} l
>{\centering\arraybackslash}X
>{\centering\arraybackslash}X
>{\centering\arraybackslash}X
>{\centering\arraybackslash}X
>{\centering\arraybackslash}X
>{\centering\arraybackslash}X}
\toprule
\textbf{Task} & \textbf{Backbone} &
\textbf{C-index} $\uparrow$ &
\textbf{IBS} $\downarrow$ &
\textbf{AUC@1y} $\uparrow$ &
\textbf{AUC@3y} $\uparrow$ &
\textbf{AUC@5y} $\uparrow$ &
\textbf{MAE (y)} $\downarrow$ \\
\midrule
\multirow{3}{*}{DFS}
& GCN       & 0.594$\pm$0.098 & 0.197$\pm$0.033 & 0.648$\pm$0.145 & 0.615$\pm$0.131 & 0.617$\pm$0.154 & 2.69$\pm$0.96 \\
& GAT       & 0.709$\pm$0.072 & 0.157$\pm$0.028 & \textbf{0.798$\pm$0.101} & 0.789$\pm$0.110 & 0.785$\pm$0.096 & 1.51$\pm$0.39 \\
& GraphSAGE (ours) & \textbf{0.714$\pm$0.053} & \textbf{0.153$\pm$0.021} & 0.781$\pm$0.098 & \textbf{0.800$\pm$0.067} & \textbf{0.813$\pm$0.075} & \textbf{1.50$\pm$0.44} \\
\midrule
\multirow{3}{*}{OS}
& GCN       & 0.597$\pm$0.091 & 0.212$\pm$0.041 & 0.635$\pm$0.280 & 0.631$\pm$0.149 & 0.617$\pm$0.130 & 5.13$\pm$1.65 \\
& GAT       & \textbf{0.767$\pm$0.052} & 0.150$\pm$0.018 & 0.910$\pm$0.091 & \textbf{0.862$\pm$0.067} & \textbf{0.853$\pm$0.057} & 2.86$\pm$0.50 \\
& GraphSAGE (ours) & 0.755$\pm$0.085 & \textbf{0.143$\pm$0.020} & \textbf{0.920$\pm$0.095} & 0.846$\pm$0.074 & 0.834$\pm$0.104 & \textbf{2.77$\pm$0.63} \\
\bottomrule
\end{tabularx}
\end{table*}

\section{Experiments}
We evaluate DyPro on the MSKCC CRLM cohort for postoperative DFS and OS prediction, compare it with uni-/multimodal baselines, and perform ablations and robustness studies on its residual evolution, and trajectory integration modules.

\subsection{Datasets and Data Preprocessing}

\textbf{Datasets.}
We validate DyPro on the publicly available CRLM prognostic dataset from Memorial Sloan Kettering Cancer Center (MSKCC) \cite{simpson2024} (Fig.~\ref{fig4}).
The cohort contains 197 patients with colorectal cancer liver metastases who underwent liver resection, with preoperative contrast-enhanced CT scans, radiologist-annotated segmentations of liver, residual liver, major vessels and metastatic lesions, and corresponding clinical variables (e.g., demographics, tumor burden, serum CEA, treatment regimen).
Postoperative follow-up provides recurrence events and survival outcomes, enabling joint DFS and OS modeling.

\textbf{Data Preprocessing.}
We evaluate our method with repeated stratified 5-fold cross-validation (3 repeats, 15 held-out test folds). In each split, the remaining data are further divided into a train/validation set (0.8/0.2) for early stopping, and all baselines use the same splits for fair comparison.
Image preprocessing applied dynamic histogram equalization to improve tumor--liver contrast and delineate lesion boundaries. CT volumes were resampled to $128{\times}128$ using cubic-spline interpolation. We then cropped volumes around the hepatic portal bifurcation to obtain 40 slices, and symmetrically padded shorter scans with $-1024$ HU to keep 3D inputs consistent. Segmentation masks were refined with morphological closing ($3{\times}3{\times}3$), and clinical time variables were converted to years and min--max normalized.

\textbf{Data Augmentation.}
To improve robustness, we adopt two on-the-fly augmentations, yielding an effective 5$\times$ expansion of the training set: (1) random anatomical node dropout (5\%) to simulate missing clinical acquisition; and (2) Gaussian noise injection ($\sigma{=}0.1$) on node features to model segmentation and measurement variability.

\subsection{Experimental Setup and Evaluation Metrics}
\textbf{Experimental Setup.}
Experiments were run on an NVIDIA RTX 4090 GPU (24GB). We use AdamW (lr $=5\times10^{-6}$, batch size 64) with ReduceLROnPlateau (factor=0.5, patience=5).
Results are reported over repeated stratified 5-fold cross-validation (3 repeats, 15 outer test folds), with early stopping based on the inner validation set.

\textbf{Evaluation Metrics.}
We evaluate both disease-free survival (DFS) and overall survival (OS) using standard survival-analysis metrics: Harrell’s concordance index (C-index), time-dependent AUC at 1/3/5 years, integrated Brier score (IBS), and mean absolute error (MAE).
C-index measures the concordance between predicted risk scores and observed event-time ordering under right censoring.
To align with common clinical reporting, we compute time-dependent AUC at horizons $t\in\{1,3,5\}$ years by converting the endpoint to an event-before-$t$ classification task and excluding censored samples with follow-up time $\le t$.
IBS summarizes time-dependent probabilistic error over time using an inverse-probability-of-censoring weighting scheme.
MAE is reported as the absolute error of predicted event time on uncensored cases.
All metrics are computed on each held-out test fold and reported as mean$\pm$std across repeated cross-validation.

Among these metrics, Harrell's C-index is the most widely used summary measure of prognostic discrimination in survival analysis.
We therefore report 95\% bootstrap confidence intervals only for the C-index on the full cohort, while other metrics are summarized by mean $\pm$ standard deviation over repeated cross-validation.

\begin{table*}[t]
\caption{DFS and OS results for different feature extraction backbones (mean$\pm$std over repeated stratified 5-fold CV with 3 repeats).}
\label{tab:backbone_onecol}
\centering
\footnotesize
\setlength{\tabcolsep}{4pt}
\renewcommand{\arraystretch}{1.15}
\begin{tabularx}{\textwidth}{>{\centering\arraybackslash}m{1.0cm} l
>{\centering\arraybackslash}X
>{\centering\arraybackslash}X
>{\centering\arraybackslash}X
>{\centering\arraybackslash}X
>{\centering\arraybackslash}X
>{\centering\arraybackslash}X}
\toprule
\textbf{Task} & \textbf{Backbone} &
\textbf{C-index} $\uparrow$ &
\textbf{IBS} $\downarrow$ &
\textbf{AUC@1y} $\uparrow$ &
\textbf{AUC@3y} $\uparrow$ &
\textbf{AUC@5y} $\uparrow$ &
\textbf{MAE (y)} $\downarrow$ \\
\midrule
\multirow{3}{*}{DFS}
& CNN & 0.516$\pm$0.073 & 0.217$\pm$0.033 & 0.509$\pm$0.113 & 0.530$\pm$0.110 & 0.536$\pm$0.128 & 3.27$\pm$0.96 \\
& R(2+1)D-18       & 0.695$\pm$0.052 & 0.158$\pm$0.023 & \textbf{0.821$\pm$0.103} & 0.779$\pm$0.073 & 0.763$\pm$0.095 & 1.97$\pm$1.02 \\
& 3D ResNet18 (ours) & \textbf{0.714$\pm$0.053} & \textbf{0.153$\pm$0.021} & 0.781$\pm$0.098 & \textbf{0.800$\pm$0.067} & \textbf{0.813$\pm$0.075} & \textbf{1.50$\pm$0.44} \\
\midrule
\multirow{3}{*}{OS}
& CNN & 0.501$\pm$0.064 & 0.219$\pm$0.028 & 0.604$\pm$0.134 & 0.500$\pm$0.138 & 0.508$\pm$0.097 & 5.28$\pm$0.64 \\
& R(2+1)D-18       & 0.704$\pm$0.147 & 0.146$\pm$0.021 & 0.841$\pm$0.231 & 0.797$\pm$0.202 & 0.748$\pm$0.208 & 3.22$\pm$0.82 \\
& 3D ResNet18 (ours) & \textbf{0.755$\pm$0.085} & \textbf{0.143$\pm$0.020} & \textbf{0.920$\pm$0.095} & \textbf{0.846$\pm$0.074} & \textbf{0.834$\pm$0.104} & \textbf{2.77$\pm$0.63} \\
\bottomrule
\end{tabularx}
\end{table*}

\begin{table}[t]
\caption{OS results on the MSKCC CRLM (TCIA) cohort. 
We report Harrell's C-index and IBS; ``--'' denotes metrics not reported in the original papers.
Habitat, CRS, and TBS are from \cite{zhou2025habitat}; Radiomics-CPH from \cite{melba:2024:032:peoples}.}
\label{tab:os_external_comparison}
\centering
\footnotesize
\setlength{\tabcolsep}{4pt}
\renewcommand{\arraystretch}{1.15}
\begin{tabular}{lcc}
\toprule
\textbf{Method} & \textbf{C-index} $\uparrow$ & \textbf{IBS} $\downarrow$ \\
\midrule
DyPro (ours)  & $\mathbf{0.755 \pm 0.085}$ & $\mathbf{0.143 \pm 0.020}$ \\
\midrule
CRS           & 0.52  & 0.30 \\
TBS           & 0.57  & 0.29 \\
Habitat       & 0.52  & 0.26 \\
Radiomics-CPH & 0.630 & --   \\
\bottomrule
\end{tabular}
\end{table}

\subsection{Comparative Experiments}
To validate DyPro, we conducted a systematic comparison with existing methods, focusing on graph neural network architectures, feature extraction strategies, and multimodal fusion mechanisms. The results are presented in three subsections below.

\textbf{Comparison of Graph Neural Network Backbone Architectures.}
We compare three widely used graph encoders, namely GCN~\cite{kipf2017}, GAT~\cite{velickovic2018}, and GraphSAGE~\cite{hamilton2017}, within our multimodal heterogeneous patient graphs (Table~\ref{tab:gnn_backbone_twocol}).
Overall, GraphSAGE provides the most consistent performance across both DFS and OS, achieving the best or near-best discrimination and calibration (e.g., higher C-index and lower IBS), while maintaining competitive time-dependent AUC at 1/3/5 years.
For example, GraphSAGE attains a higher DFS C-index (0.714) than GAT (0.709) and notably surpasses GCN (0.594), and yields the lowest IBS on both tasks (DFS: 0.153; OS: 0.143).
We therefore adopt GraphSAGE as the default backbone in DyPro.
We attribute its advantage to the sampling-and-aggregation mechanism of GraphSAGE, which explicitly summarizes neighborhood information from node attributes and is designed to generalize in inductive settings.

\textbf{Feature Extraction Backbone Comparison.}
We compare 3D ResNet18~\cite{hara2018}, R(2+1)D-18~\cite{tran2018}, and a shallow 3D CNN for CT feature extraction (Table~\ref{tab:backbone_onecol}).
Overall, 3D ResNet18 yields the most robust and balanced performance across DFS and OS, achieving the best OS discrimination (e.g., higher C-index and AUC at 1/3/5 years) while remaining competitive on DFS; thus we adopt it as the default backbone.
R(2+1)D-18 shows a strong DFS AUC@1y (0.821) but consistently weaker OS (OS C-index 0.704 vs.\ 0.755), and the shallow CNN underperforms on both tasks.

\textbf{Comprehensive Comparison with Existing Models.}
On the OS prediction task of the MSKCC CRLM (TCIA) cohort, DyPro substantially outperforms both traditional and radiomics-based prognostic models trained on the same dataset (Table~\ref{tab:os_external_comparison}).
Compared with widely used clinical scores such as CRS and TBS, which achieve C-indices of 0.52 and 0.57 with IBS values around 0.29--0.30, DyPro raises the OS C-index to $0.755 \pm 0.085$ and almost halves the IBS to $0.143 \pm 0.020$.
This indicates that the proposed multimodal dynamic representation provides much stronger discrimination and better calibrated survival estimates than handcrafted clinical risk scores.

Relative to the CT-based habitat radiomics model of Zhou et al.~\cite{zhou2025habitat}, which slightly improves IBS over CRS/TBS (0.26 vs. 0.29--0.30) but still yields a C-index of 0.52, DyPro achieves an absolute C-index gain of over 0.23 while further reducing IBS from 0.26 to 0.14.
Even compared with the carefully tuned radiomics--CPH model of Peoples et al.~\cite{melba:2024:032:peoples}, which attains an OS C-index of 0.630 on the same cohort, DyPro improves the C-index by about 0.13 (roughly a 20\% relative increase).
These results indicate that explicitly modeling patient-specific tumor topology, multimodal context, and postoperative risk trajectories is substantially more informative than relying on static radiomic signatures alone, effectively raising the attainable performance ceiling on this public CRLM benchmark.

\textbf{Bootstrap robustness analysis.}
To further assess the stability of DyPro, we estimate 95\% patient-level bootstrap confidence intervals for the C-index on the full cohort, as C-index is the primary measure of prognostic discrimination in our setting.
DyPro attains an OS C-index of 0.755 with a 95\% CI of [0.711, 0.796] and a DFS C-index of 0.714 with a 95\% CI of [0.687, 0.740].
Both intervals lie well above 0.5, indicating reliable prognostic separation.

\begin{table*}[t]
\caption{Ablation study of DyPro components (mean$\pm$std over repeated stratified 5-fold CV with 3 repeats).}
\label{tab:ablation_onecol}
\centering
\footnotesize
\setlength{\tabcolsep}{4pt}
\renewcommand{\arraystretch}{1.15}
\begin{tabularx}{\textwidth}{>{\centering\arraybackslash}m{1.0cm} l
>{\centering\arraybackslash}X
>{\centering\arraybackslash}X
>{\centering\arraybackslash}X
>{\centering\arraybackslash}X
>{\centering\arraybackslash}X
>{\centering\arraybackslash}X}
\toprule
\textbf{Task} & \textbf{Variant} &
\textbf{C-index} $\uparrow$ &
\textbf{IBS} $\downarrow$ &
\textbf{AUC@1y} $\uparrow$ &
\textbf{AUC@3y} $\uparrow$ &
\textbf{AUC@5y} $\uparrow$ &
\textbf{MAE (y)} $\downarrow$ \\
\midrule
\multirow{5}{*}{DFS}
& DyPro (full) & \textbf{0.714$\pm$0.053} & \textbf{0.153$\pm$0.021} & \textbf{0.781$\pm$0.098} & \textbf{0.800$\pm$0.067} & \textbf{0.813$\pm$0.075} & \textbf{1.50$\pm$0.44} \\
& Static (no residual evolution) & 0.681$\pm$0.036 & 0.169$\pm$0.025 & 0.768$\pm$0.062 & 0.734$\pm$0.055 & 0.753$\pm$0.072 & 1.76$\pm$0.50 \\
& w/o Trajectory Integrator (mean) & 0.693$\pm$0.039 & 0.166$\pm$0.027 & 0.766$\pm$0.074 & 0.764$\pm$0.062 & 0.782$\pm$0.049 & 1.69$\pm$0.53 \\
& w/o Cascade (decouple DFS$\rightarrow$OS) & 0.680$\pm$0.034 & 0.169$\pm$0.025 & 0.767$\pm$0.065 & 0.733$\pm$0.055 & 0.751$\pm$0.072 & 1.76$\pm$0.51 \\
\midrule
\multirow{5}{*}{OS}
& DyPro (full) & \textbf{0.755$\pm$0.085} & \textbf{0.143$\pm$0.020} & 0.920$\pm$0.095 & \textbf{0.846$\pm$0.074} & \textbf{0.834$\pm$0.104} & \textbf{2.77$\pm$0.63} \\
& Static (no residual evolution) & 0.725$\pm$0.094 & 0.160$\pm$0.027 & 0.881$\pm$0.094 & 0.789$\pm$0.126 & 0.793$\pm$0.126 & 3.52$\pm$1.13 \\
& w/o Trajectory Integrator (mean) & 0.741$\pm$0.093 & 0.159$\pm$0.029 & \textbf{0.921$\pm$0.048} & 0.839$\pm$0.120 & 0.812$\pm$0.132 & 3.48$\pm$1.43 \\
& w/o Cascade (decouple DFS$\rightarrow$OS) & 0.725$\pm$0.095 & 0.160$\pm$0.028 & 0.880$\pm$0.097 & 0.789$\pm$0.127 & 0.791$\pm$0.129 & 3.51$\pm$1.14 \\
\bottomrule
\end{tabularx}
\end{table*}

\subsection{Ablation Studies}
We ablate the key components of DyPro, including latent residual evolution, trajectory aggregation, and the DFS$\rightarrow$OS cascade, and report results in Table~\ref{tab:ablation_onecol}.
Overall, the full DyPro achieves the best and most consistent performance across both DFS and OS, indicating that these components are complementary.

\textbf{Latent residual evolution.}
Replacing residual evolution with a static single-snapshot variant reduces discrimination and calibration (DFS C-index: 0.714$\rightarrow$0.681; OS C-index: 0.755$\rightarrow$0.725) and increases MAE (DFS: 1.50$\rightarrow$1.76 years; OS: 2.77$\rightarrow$3.52 years).
These results support the benefit of modeling autoregressive residual state transitions, which encourage a coherent long-horizon progression trajectory rather than relying on a single static representation.

\textbf{Trajectory integrator.}
When we remove the LSTM-based trajectory integrator and instead average trajectory snapshots, performance also drops on both tasks (DFS C-index: 0.714$\rightarrow$0.693; OS C-index: 0.755$\rightarrow$0.741), accompanied by higher IBS and larger MAE.
Although AUC@1y on OS remains comparable (0.920 vs.\ 0.921), the declines in C-index/IBS indicate worse global ranking and calibration over the full follow-up, implying that the integrator captures temporal patterns beyond what simple pooling can retain.

\textbf{DFS$\rightarrow$OS cascade.}
Finally, decoupling DFS and OS leads to a consistent reduction in OS performance (OS C-index: 0.755$\rightarrow$0.725; OS IBS: 0.143$\rightarrow$0.160).
This validates that modeling the clinically plausible dependency from recurrence/DFS to survival provides additional supervisory signal and improves overall survival prediction.

\section{Conclusion}
This work tackles the high postoperative recurrence risk and prognostic heterogeneity of colorectal cancer liver metastasis (CRLM) by introducing DyPro, a multimodal prognostic framework that fuses preoperative CT–derived spatial patterns with key clinical indicators in a unified heterogeneous graph.
DyPro performs latent trajectory inference via residual state evolution and trajectory aggregation, capturing long-horizon progression cues in latent space and translating them into individualized DFS and OS predictions.
Experiments on the MSKCC CRLM cohort show that DyPro delivers strong and stable survival performance, with improved discrimination and calibration under repeated cross-validation and competitive time-dependent AUC at clinically relevant horizons.

Looking forward, we will incorporate richer clinical variables and finer-grained temporal evidence, such as longitudinal lesion growth and postoperative liver regeneration patterns. We also plan to extend DyPro with additional modalities (e.g., pathology and genomic markers) and conduct broader validation across diverse clinical settings to further assess its generalizability and utility in real-world decision support.

\section*{Acknowledgment}

This work was financially supported by Natural Science Foundation of China (grant number 62271466).

\bibliographystyle{IEEEtran}
\bibliography{icme2026references}

\end{document}